\documentclass[fleqn,10pt]{wlscirep}
\usepackage{subfig}
\usepackage{dsfont}
\usepackage{mathtools}
\usepackage{epstopdf} 

\title{Squeezing-enhanced quantum key distribution over atmospheric channels}

\author[1,*]{Ivan Derkach}
\author[1,$\dagger$]{Vladyslav C. Usenko}
\author[1,$\ddagger$]{Radim Filip}
\affil[1]{Department of Optics, Palacky University, 17. listopadu 50,  772~07 Olomouc, Czech Republic}

\affil[*]{ivan.derkach@upol.cz}
\affil[$\dagger$]{usenko@optics.upol.cz}
\affil[$\ddagger$]{filip@optics.upol.cz}

\begin{abstract}
We propose the Gaussian continuous-variable quantum key distribution using squeezed states in the composite channels including atmospheric propagation with transmittance fluctuations. We show that adjustments of signal modulation and use of optimal feasible squeezing can be sufficient to significantly overcome the coherent-state protocol and drastically improve the performance of quantum key distribution in atmospheric channels, also in the presence of additional attenuating and noisy channels. Furthermore, we consider examples of atmospheric links of different lengths, and show that optimization of both squeezing and modulation is crucial for reduction of protocol downtime and increase of secure atmospheric channel distance. Our results demonstrate unexpected advantage of fragile squeezed states of light in the free-space quantum key distribution applicable in daylight and stable against atmospheric turbulence.
\end{abstract}
\begin{document}

\flushbottom
\maketitle
\thispagestyle{empty}

\section*{Introduction}

Quantum key distribution (QKD) \cite{Gisin02, Scarani09, Diamanti2016} is one of the major practical applications of quantum information theory, which provides trusted parties (Alice and Bob) with the methods (protocols) for provably secure distribution of secret cryptographic keys so that security of the key can be verified using fundamental principles of quantum physics. One of the main requirements of QKD is the availability of a dedicated quantum channel capable of transmitting coherent quantum signals between the sending and receiving stations. In the case of fiber-optical channels, being the typical media for QKD implementations, this means a dedicated optical fiber, possibly with co-existing classical or quantum signals. However, the dedicated fiber-optical infrastructure can be unavailable, e.g., in the case of movable stations, necessity of quick channel deployment or in hostile environments. Moreover, the extra-long-distance inter-continental quantum communication over satellites relies on the free-space channels \cite{Bedington2017}. Therefore, the free-space channels are an important physical medium for QKD implementations.

The main issue faced by the discrete-variable (DV) QKD protocols, based on single-photon states or weak coherent pulses and the direct photon counting, is the sensitivity of the detectors to the background light, which adds noise to the measured data. This renders standard DV QKD protocols practically unusable in the daylight conditions unless spectral filtering is applied, which adds unwanted additional loss and complexity to the set-up. At the same time, applicability and efficiency are crucial for QKD as they directly affect the secret communication, based on the quantum-secure keys. Alternatively, continuous-variable (CV) QKD protocols \cite{Braunstein2005, Pirandola2008, Weedbrook2012, Diamanti15}, based on the multiphoton coherent \cite{Grosshans2003,Lodewyck2007,Fossier2009,Jouguet:12,Huang16} or squeezed states \cite{Cerf01} and homodyne quadrature detection using off-the-shelf equipment, can overcome this limitation. Indeed, a homodyne detector, which matches a signal to a narrow-band local oscillator (LO) beam, being the phase reference for the measurement, can intrinsically filter out the background radiation and make CV QKD protocols directly applicable in the daylight. However, CV QKD protocols are known to be sensitive to transmittance fluctuations, caused by the atmospheric turbulence\cite{Berman06, Semenov09}. Such fluctuations, also referred to as the channel fading, result in the excess noise, which is proportional to the quadrature variance of a signal beam and limits applicability of CV QKD over atmospheric channels\cite{Usenko2012}, which is also valid for the recently studied fast-fading channels\cite{Papanastasiou2018}. It was shown that noise due to the channel fading leads to security break of coherent-state CV QKD protocol and requires optimization of modulation and sub-channel post-selection for long-distance implementations over turbulent channels\cite{Usenko2012}. As a possible alternative solution, the use of squeezed signal states in CV QKD can be considered. Indeed, the squeezed states are known to be more robust against CV QKD imperfections such as inefficient post-processing\cite{vmo2} or strong channel excess noise\cite{GP09, Madsen2012}.

In the current paper we suggest squeezed-state protocol for free-space CV QKD with the channel fading and study the applicability and robustness of the protocol to realistic imperfections in comparison to the coherent-state protocol. We confirm the positive effect of signal state squeezing in realistic free-space CV QKD, taking into account channel fluctuations and additional fixed losses as well as other practical imperfections, such as limited post-processing efficiency. We show that squeezing can be helpful in the fluctuating channels, but should be optimized for the given conditions together with modulation used to encode information. We verify the results by considering the fading channel model, based on the beam wander, which is the dominating effect, causing free-space quantum channel fluctuations\cite{Vasylyev2012, Vasylyev2016}. Furthermore, we confirm the advantage of squeezed-state protocol using characteristics of the real atmospheric channels and show that the use of squeezed signals allows extending the secure distance of the CV QKD protocols. The advantage is stable against the finite-size effects of limited data ensembles and impurity of the squeezed signal states. Our results therefore pave the way for efficient free-space QKD realization in daylight conditions, robust against atmospheric turbulence effects. 

The paper is structured as follows. In Sec. \ref{Protocol} we describe used CV QKD scheme that allows to independently control and manipulate squeezing and displacement in both quadratures of the signal state. Sec. \ref{General} is devoted to analytical description of the effects of general fading channels on coherent- and squeezed-state protocols. Finally in Sec. \ref{simulations} we compare the performances of various CV QKD protocols in real and modeled noisy composite untrusted channels. 

\section{CV QKD protocol}\label{Protocol}

The goal of a QKD protocol is to share a correlated string of data between two trusted parties, usually referred to as Alice and Bob. To do so Alice first prepares a quantum state of light, and encodes key bits into it. In the current work we will operate in the Gaussian regime, meaning the states used are described on phase space by the Gaussian Wigner function \cite{Weedbrook2012}. In this case, we can use Gaussian approximation of the states, channel and measurement and use the powerful covariance matrix formalism to simplify analysis to the finite-dimensional case \cite{Weedbrook2012}. However, secure key rate is a complex nonlinear functional of the elements of covariance matrix, therefore a usefulness of nonclassical and entangled states has to be analyzed in detail.  Hence we suppose that Alice generates coherent or squeezed Gaussian states (using a laser source or e.g. optical parametric oscillator, respectively) and displaces them on a phase space in one or both quadratures ($X$ or/and $P$) according to two independent Gaussian distributions with zero mean and variance $V_m$. Such a scheme encodes two real numbers from a continuous Gaussian distribution and therefore it has much higher capacity per time interval than a discrete encoding. Due to recent developments in the field of quantum optics, both coherent and displaced squeezed states can be generated with sufficient purity \cite{Serikawa2016, Vahlbruch2016}. The signal state prior to modulation can be therefore described by the diagonal covariance matrix $\gamma_B=diag[V_s,1/V_s]$, where $V_s$ is the variance of the squeezed quadrature, here and further, without loss of generality, it is assumed to be the $X$ quadrature (covariance matrix for a coherent state reduces to a $2\times2$ unity matrix). Squeezed states are known to be more sensitive to a loss than coherent states, however, still squeezing never vanishes under pure loss. The signal after the modulation is characterized by the covariance matrix $\gamma_B'=diag[V_s+V_m,1/V_s+b V_m]$, where for the coherent-state protocol $b=1$ and it corresponds to modulation in both quadratures, while for the squeezed-state protocol $b=0$, which indicates that only squeezed quadrature is modulated. We omit optimization over $b$, because we focus on testing of squeezed state applicability. Displaced signal states together with LO are sent to Bob via untrusted quantum channel where the signal suffers from losses and noise (LO can be also reconstructed locally \cite{Qi2015,Soh2015,Marie2017}). An eavesdropper Eve is presumed to be the cause of both losses and noise within the channel, is able to obtain and store the information about signal states, and is limited in her attacks on the channel only by the laws of physics. Bob on his side conducts a homodyne measurement with an auxiliary LO, and proceeds to key sifting, error correction, and privacy amplification using an authenticated classical channel established beforehand with Alice. As the outcome, Bob produces a sequence of secure bits shared with Alice.\par

\begin{figure}[t] 
\centering
\subfloat[P\&M CV QKD scheme]{\includegraphics[width=0.55\textwidth]{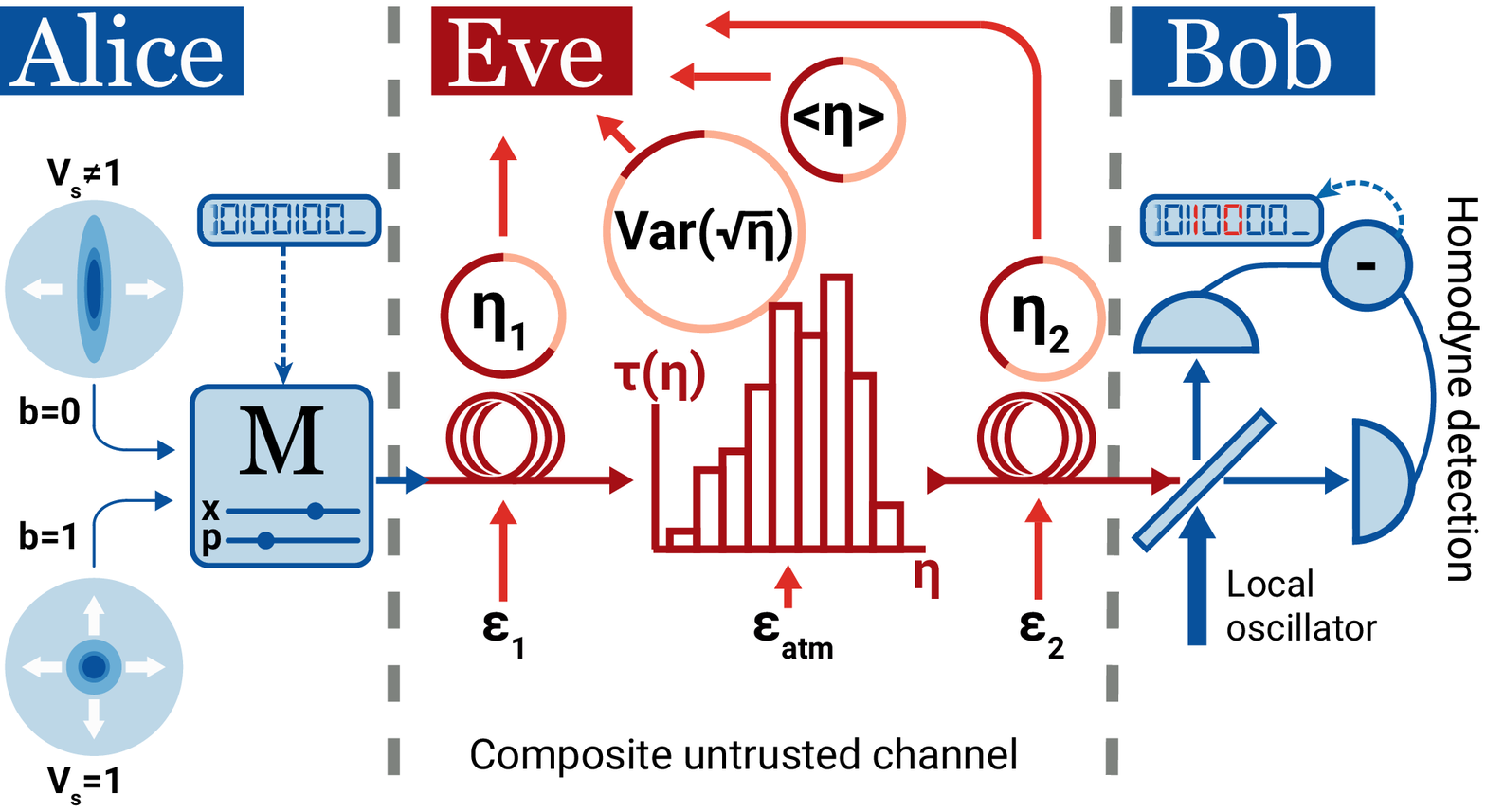}\label{PM:1}}
\subfloat[Secure key rate for squeezed state protocol]{\includegraphics[width=0.4\textwidth]{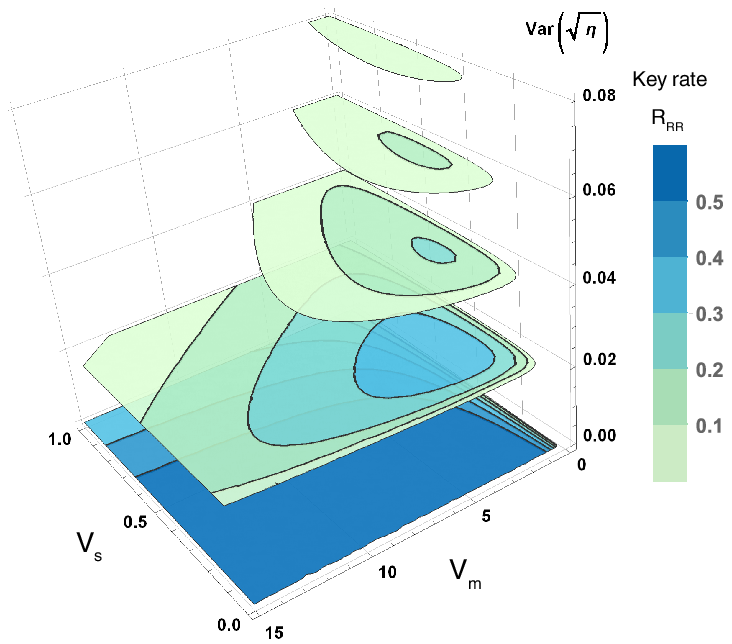}\label{PM:2}}
\caption{(\textbf{a}) Prepare-and-measure CV QKD scheme. Alice prepares a Gaussian (squeezed $V_s< 1$ or coherent $V_s=1$) state of light and applies displacement operation in one ($b=0$) or both ($b=1$) quadratures. The signal is sent though the untrusted quantum channel to Bob, where the latter conducts homodyne detection. Composite unstrusted channel can consist of noisy fiber channels with fixed losses $\eta_{1,2}$, and excess noise $\epsilon_{1,2}$, respectively, and atmospheric channel $\tau(\eta)$ [characterized by mean losses $\langle\eta\rangle$, and transmittance fluctuations strength $Var(\sqrt{\eta})$, and excess noise $\epsilon_{atm}$]. Eve can receive information from all the channels, and is considered to perform collective attacks on the protocol. (\textbf{b}) Positive key rate (in bits per channel use) for squeezing $V_s$, modulation variance $V_m$ and fading strength Var($\sqrt{\eta}$), $\langle\eta\rangle=1/2,\eta_{1}=\eta_2=1, \epsilon_{1}=\epsilon_{2}=\epsilon_{atm}=0$. The key rate values range from $0<R\leqslant0.1$ (lightly shaded areas) to $R>0.5$ (darkest shaded areas) with 0.1 step. It is evident that the increase in fluctuations strength significantly decreases the key rate, however squeezing and modulation optimization can be conducted in order to improve the performance of the protocol. Importantly, optimal performance can be reached by a feasible squeezing for nonvanishing $Var(\sqrt{\eta})$.}
    \end{figure}

The implementation of the basic coherent- or squeezed-state protocol described above is usually referred to as prepare-and-measure (P\&M), and it is depicted on Fig. \ref{PM:1}. In a realistic scenario, a short distance P\&M free-space QKD will combine fiber-based channels in the buildings before and after the flexible atmospheric channels between the buildings. To predict protocol applicability, the parts of untrusted channel are considered to be either characterized by fixed transmittance (which can correspond to the fiber-optical parts of the entire link) or by fluctuating transmittance, which most commonly correspond to free-space atmospheric links. In the current work we consider a CV QKD protocol realization in a hybrid case, when an untrusted channel may consist of a combination of channels with fixed transmittance $\eta_{1,2}$ (fiber based channels), and a free-space channel with fluctuating transmittance $\eta$, governed by a probability distribution $\tau(\eta)$, in the middle. Furthermore both kinds of channels are not restricted to pure losses, but can introduce excess noise, the latter however is assumed to be fixed throughout the duration of all key distribution. Respectively, the excess noise in fiber channels is $\epsilon_{1,2}$, and in free-space channel $\epsilon_{atm}$, while total noise added in the channel and measured by Bob is $\epsilon_{+}$.  Such the excess noise can be small in practice, however, it is important to introduce it in analysis to understand its impact.  \par

Security of a CV QKD protocol is defined in terms of positivity of the lower bound on the rate (in bits per channel use) of the secret key\cite{CohAtt, Gehring2015} distributed among the trusted parties. Either Alice or Bob must agree to be the reference side of the protocol, which means they will perform direct (DR) or reverse (RR) reconciliation\cite{Grosshans2003b, cv3}, respectively. Even though the efficiency $\beta\in[0,1]$ of the algorithms for reconciliation is close to unity, it must be accounted for when estimating the key rate of the protocol. We assume pessimistic scenario where Eve is able to purify the state shared between the trusted parties, and conduct collective measurement over her part of the state. The strategy, presumably used by Eve, is called collective attack, and the key rate\cite{Devetak2005} of the protocol under such attack can be written as: 
  \begin{equation}\label{Key}
        R_{DR}=\beta I_{AB}-\chi_{AE}, \qquad R_{RR}=\beta I_{AB}-\chi_{BE},
    \end{equation}
where $\chi_{AE(BE)}$ is the Holevo bound \cite{Holevo2001} - an upper bound on the information accessible to Eve with respect to the trusted reference side. Quantity $I_{AB}$ in Eq.\ref{Key} is the mutual information between trusted parties. For more details on the security analysis see Supplementary Information. 

\section{General fading channel influence}\label{General} 

Let us first have a look on the sole influence of the channel with transmittance fluctuations on CV QKD protocol. The atmospheric effects present in free-space link will unavoidably affect the transmitted beam, which will experience fading. Hence the transmission coefficient must be described in terms of transmittance probability distribution $\tau(\eta)$, as opposed to a fixed loss in a fiber link. It was shown \cite{Dong2010} that such fading channel can be decomposed into a set of subchannels $\{\eta_{j}\}$ - channels with negligible attenuation fluctuations within them, which occur with probability $\tau(\eta_{j})$ so that $\sum_{j=1}^{\infty} \tau(\eta_{j}) = 1$. Now the Gaussian Wigner function of a state after a fading channel is a weighed sum of Winger functions after individual subchannels associated with fixed attenuation $\eta_{j}$ \cite{Usenko2012}. In other words the resulting shared state, described by the covariance matrix $\gamma_{AB}$, is a mixture of the states $\gamma_{AB}^{j}$ after each subchannel, which within the covariance matrix formalism is expressed by averaging over fluctuating transmittance values. \par

Statistical properties of the transmittance distribution $\tau(\eta)$ that directly affect the covariance matrix of the shared state $\gamma_{AB}$ and influence the performance of the protocol (\ref{Key}), are the mean value of transmittance $\langle\eta\rangle$, and mean value of square root of transmittance $\langle\sqrt{\eta}\rangle$. They define the fading variance\cite{Dong2010} $Var(\sqrt{\eta})=\langle\eta\rangle - \langle\sqrt{\eta}\rangle^2$. The key rate (\ref{Key}) of the protocol over a fading channel is a function of all parameters of the protocol and the channel $R(V_s, V_m, \langle\eta\rangle,Var(\sqrt{\eta}), \epsilon_{atm})$. Alternatively the overall state after a fading channel can be represented as a state after a channel with fixed attenuation \cite{Usenko2012} $\langle\sqrt{\eta}\rangle^2$, and additional variance-dependent excess noise $\epsilon_f(\tau(\eta),V_s,V_m)=Var(\sqrt{\eta})(V_s+V_m-1)$ (fixed channel excess noise $\epsilon_{atm}$ remains the same), so that $R(V_s, V_m, \langle\sqrt{\eta}\rangle^2, \epsilon_f, \epsilon_{atm})$. Protocols that use DR or RR have the same dependency on the fading $Var(\sqrt{\eta})$, with DR being limited to low attenuation channels $\langle\eta\rangle>1/2$, therefore we focus further on the reverse reconciliation since it allows one to analyze a wider range of channels, bearing in mind that the developed methodology is applicable to the protocols with DR as well.\par

In order to study solely the influence of a fading channel (for $\eta_{1,2}=1$) on the security of the CV QKD protocols, we look at the case of collective attacks conducted in case of noiseless channel ($\epsilon_{atm}=0$) and perfect post-processing accessible to trusted parties ($\beta=1$, which is a theoretical limit, but recent protocols \cite{Leverrier2008, Gehring2015, Hirano2017} are very close to it). Figure \ref{PM:2} depicts a positive key rate and it's dependency on the squeezing $V_s$ and modulation variance $V_m$ for various values of transmittance fluctuations $Var(\sqrt{\eta})$ for $\langle\eta\rangle=1/2$. The brightest colored area includes the key rate values of $R_{RR}\in (0,0.1)$, and for each consequent darker area the key rate is increased by 0.1, with the darkest area containing values $R_{RR}>0.5$. Whenever transmittance fluctuations are absent in the channel ($Var(\sqrt{\eta})=0$), stronger squeezing $V_s$ is always more beneficial for Alice and Bob. Strong squeezing allows protocols to achieve high key rate, and tolerate more losses and excess noise in the channel \cite{Madsen2012}. Typical values of fading in atmospheric channel with weak turbulence are $Var(\sqrt{\eta})\leq0.01$, while under strong turbulence one can expect at least $Var(\sqrt{\eta})>0.04$, where the uniformly distributed channel has  $Var(\sqrt{\eta})= 0.055$. The values of $Var(\sqrt{\eta})> 0.055$ correspond to a transmittance distribution described by a convex function which in the limit $Var(\sqrt{\eta})=0.25$ represents the channel with equal probabilities to either perfectly transmit the signal or fail. \par 

\begin{figure}[t] 
	\centering

\subfloat[The effect of losses and squeezing on the secure key rate]{\includegraphics[width=0.45\textwidth]{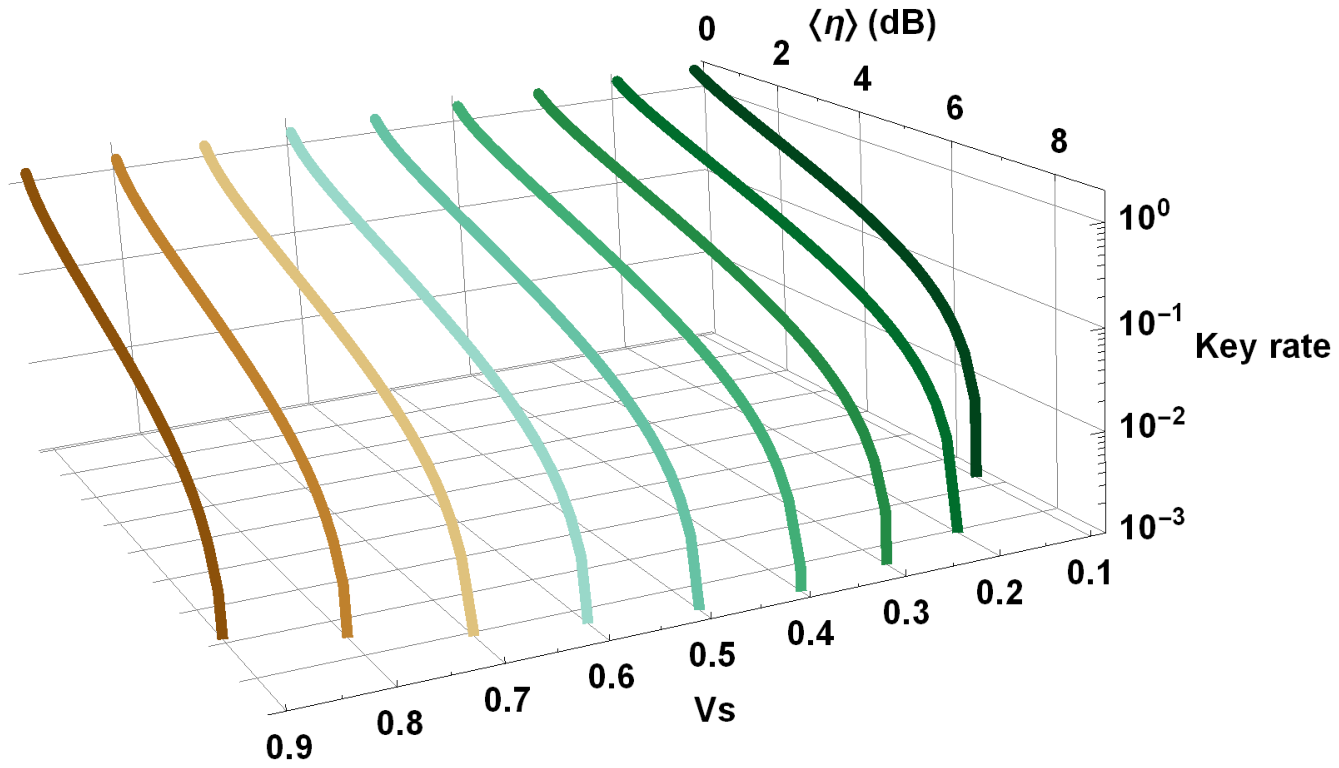}\label{FIG2:1}}
\subfloat[Optimized secure key rate in atmospheric channel]{\includegraphics[width=0.49\textwidth]{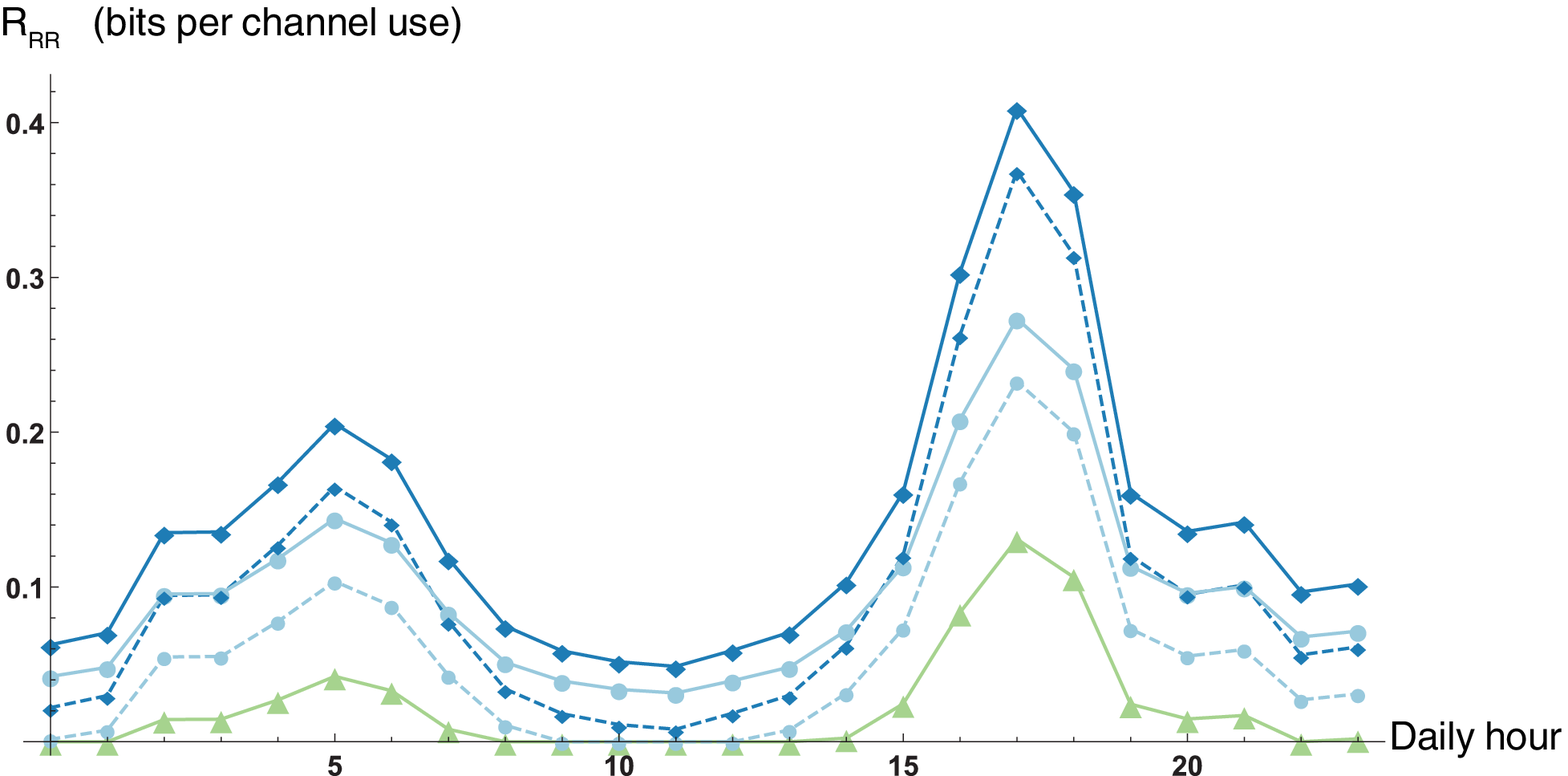}\label{FIG2:2}}

    \caption{(\textbf{a}) Secure key rate (in bits per channel use) dependency on channel losses $\langle\eta\rangle$ ($dB$) for various values of squeezing $V_s$ ($\eta_{1,2}=1$). Transmittance fluctuations are only moderate $Var(\sqrt{\eta})=0.01$, and modulation $V_m$ is optimized. Evidently high squeezing levels of carrier states in CV QKD do not necessarily translate into high tolerance to $\langle\eta\rangle$ with transmittance fluctuations. Squeezing $V_s$ and modulation variance $V_m$ should be optimized in order to sustain the most amount of losses in the untrusted channel. (\textbf{b}) Optimized secure key rate of the coherent-state protocol (green triangles), and the squeezed-state protocol (blue circles, and blue squares) in 2.2 km long atmospheric channel. Each point was obtained for the transmittance distribution simulated using the averaged (over 4 months period) hourly statistics of structure constant of refractive index of air $C_{n}^2$. Finite-size effects are considered for block sizes of $n=10^6$ (dashed lines) and $n=10^{10}$ (solid lines). The squeezed-state protocol has been optimized over both modulation $V_m$ and squeezing $V_s$, with the upper limit on the latter $V_s^{max}=-3dB$ (circles), $V_s^{max}=-10dB$ (squares). Excess noise $\epsilon_+=1\%$, efficiency $\beta=95\%$, additional losses $\eta_{comb}=-2.2dB$. The coherent-state protocol can be successfully implemented only during suitable atmospheric conditions (around 5am and 5pm). The squeezed-state protocol with limited squeezing $V_s^{max}=-3dB$ and smaller block size can on average be used during the whole day with an exception of possible temporary signal loss around 10am. However increasing the block size and/or maximal squeezing $V_s^{max}$ would have allowed to operate over such short atmospheric link continuously throughout the period of all 4 months.}
    \end{figure}

The presence of even small fluctuations of transmittance surprisingly limits applicable values of squeezing $V_s$, i.e. signal states with strong squeezing $V_s<0.02$ can render the squeezed-state protocol insecure. Such sensitivity of strongly squeezed states to transmittance fluctuations is not exhibited in individual attacks, and manifests only for the collective attacks in the Holevo bound $\chi$. It is an example why the collective effects are important to be analyzed. With the increase of transmittance fluctuations $Var(\sqrt{\eta})$ the optimal values of squeezing are shifted towards lower values, corresponding to stronger squeezing. Therefore, optimization of the squeezing with respect to the channel and information encoding by coherent modulation is required. On the other hand, modulation $V_m$ is shown to be bounded as well, and even more so, if one would take into account limited post-processing efficiency $\beta$. The need for squeezing optimization is further stressed in the fading channel where mean losses $\langle\eta\rangle$ are increasing, as is visible in the Fig. \ref{FIG2:1}. In this case the modulation $V_m$ is already optimized, and the strength of transmittance fluctuations in the fading channel is fixed to a low value $Var(\sqrt{\eta})=0.01$, however this already alters the performance of the squeezed-state protocols. Even in the presence of such minor transmittance fluctuations the protocol that uses strongly squeezed signal states (as $V_s=0.1$) cannot tolerate more losses than the one that uses significantly weaker squeezed states (as $V_s=0.9$). All these results demonstrate that a feasible squeezing is fairly sufficient for multiple increase of the secure key rate over the atmospheric channels. \par

The main takeaway of the analysis is that for a given fading channel with estimated values of $\langle\eta\rangle$ and $Var(\sqrt{\eta})$, one should optimize both squeezing $V_s$ and modulation $V_m$ in order to operate in secure regime, in the first place, and subsequently to maximize the secure key rate. In a standard entanglement-based scheme\cite{Madsen2012} the effective modulation variance $V_m$ of encoding alphabet and conditional squeezing $V_s$ are inherently connected ($V_s=1/V$ and $V_m=V-1/V$, respectively). In Fig. \ref{PM:2} the key rate will then occupy a curved plane perpendicular to $V_s-V_m$ plane, and it crosses all regions of the maximal key rate for all values of $Var(\sqrt{\eta})$. In other words, entanglement-based protocol optimized in terms of key rate will yield roughly same performance as the protocol where squeezing $V_s$ and modulation $V_m$ are optimized separately. However the P\&M protocol is simpler for experimental implementation and more flexible, meaning it can achieve the same key rate as entanglement-based protocol, with less squeezing, and compensate by applying modulation with higher variance. On the other hand, the entanglement-based protocol can be extended to a secure communication network. It is certainly stimulating for further analysis and experimental development. \par

\section{Atmospheric channel fluctuations}\label{simulations}
	In the following section we analyze the performance of both the coherent-state and the squeezed-state protocols established over a composite untrusted channel, as depicted on Fig. \ref{PM:1}. We consider limited post-processing $\beta<1$, additional fixed losses before and after the fading channel $\eta_{1,2}< 1$, as well as thermal excess noise in all channels $\epsilon_{1,2,atm}>0$. Finite-size effects \cite{Leverrier2010} were also taken into account as a correction to the key rate (\ref{Key}) $\Delta(n)$ that strongly depends on the total size $n$ of data sets shared by trusted parties. \par
	The covariance matrix describing the state after composite untrusted channel and received by Bob is $\gamma_{B}'=(\gamma_B-\mathds{1})\langle\eta\rangle\eta_{comb}+(1+\epsilon_{+})\mathds{1}$, where $\eta_{comb}=\eta_1\eta_2$ is product of all transmittance values of channels with fixed losses, and $\epsilon_{+}=\epsilon_2+\epsilon_{atm} \eta_2+\epsilon_1 \eta_2 \langle\eta\rangle$ is a total excess noise received by Bob. Even though Alice and Bob may not be able to distinguish, and properly attribute losses and noise to each individual channel, they are only required to estimate each time the overall loss $\eta_{comb}\eta_j$, and total excess noise $\epsilon_{+}$ imposed on the state that arrives to the Bob's side. 
	
\subsection{Fading model}\label{Model}
        To simulate the transmittance in free-space optical links with dissimilar properties we adopt an atmospheric transmittance probability distribution with an elliptic-beam approximation \cite{Semenov2009, Vasylyev2012, Vasylyev2016, Bohmann2016, Vasylyev2017, Bohmann2017}. The model assumes a Gaussian optical beam propagating through atmospheric horizontal link with isotropic turbulence, where the beam is distorted and suffers from broadening, deformation of beam spot into elliptical shape, as well as beam wandering. The model has been successfully applied in the regimes of weak, weak-to-moderate and strong turbulence. Furthermore, the incorporated model can be used in conjunction with experimentally employed beam tracking techniques. The probability distribution of the transmittance (PDT) is given as: 

\begin{equation}
\tau(x_0,y_0,W_0,\Theta_1,\Theta_2,\phi)= \frac{2}{\pi}\int_{\mathds{R}^4}d^4v\int^{\pi/2}_{0} d\phi \rho_G (v;\mu;\Sigma)\delta[\eta-\eta(v,\phi)].
\label{PDT}
\end{equation}

The probability (\ref{PDT}) is governed by five real parameters, that are (with an exception of $W_0$) randomly changed by the atmosphere: current position of beam-spot center $x_0, y_0$; relation of initial beam-spot radius $(W_0)$ to elliptic beam-spot semiaxes $(W_1, W_2)$ - $\Theta_1,\Theta_2$; uniformly distributed angle of semiaxis of elliptical beam-spot relative to $x$-axis $\phi$. Transmittance $\eta(v,\phi)$ in (\ref{PDT}) is defined as:

\begin{equation}
\eta=\eta_0\exp\Biggl\{-\Bigg[\frac{r_0/a}{R\left(\frac{2}{W_{eff}(\phi-\varphi_0)}\right)} \Bigg]^{\lambda\big( \frac{2}{W_{eff}(\phi-\varphi_0)}\big)} \Biggl\},
\label{Eta}
\end{equation}

where $\eta_0$ is transmittance of the beam in the center of aperture ($x_0=y_0=0$), $R(\xi), \lambda(\xi)$ are scale and shape functions, effective spot radius of the circular beam $W_{eff}$, and aperture radius $a$. Vector $r_0=(x_0,y_0,\varphi_0)^T$ is a beam spot center deviation from aperture center. Vector $v$ with mean $\mu$ and covariance matrix $\Sigma$ regulate $\rho_G(v;\mu;\Sigma)$ from Eq.(\ref{PDT}).\par

	We assume perfect channel estimation and rely on assessment of parameters\cite{Vasylyev2016} $v,\mu,\Sigma$  in regimes of weak,weak-to-moderate and strong turbulence and carry out an atmospheric channel transmittance simulation in respective regimes by Monte Carlo method. The simulation of transmittance (\ref{Eta}) in appropriate turbulence regime is ultimately driven by beam wave-number $k$, propagation distance $L$ and $C^2_n$ structure constant of the refractive index of the air, as well as initial beam-spot radius $W_0$. An important description of atmospheric links that incorporates most of aforementioned quantities is the Rytov parameter $\sigma^2_R=1.23C^2_nk^{\frac{7}{6}}L^{\frac{11}{6}}$. When the atmospheric turbulence is considered to be weak or weak-to-moderate the dominating effect is the beam wandering, while beam broadening and deformation is minor, the Rytov parameter takes values up to $\sigma_R^2\lesssim1$. The regime when both beam wandering and deformation effects are present and non-negligible corresponds to strong turbulence and the Rytov parameter values are $\sigma_R^2\gg1$.

\subsection{Analysis of real channels}\label{Example}

The first example we consider is a CV QKD protocol established over an atmospheric channel of fixed length of 2.2 $km$ that is used for an extensive period of time. The transmittance values have been simulated, using the elliptic-beam model (\ref{Eta}), based on the data obtained by Czech Metrology Institute from atmospheric channel in the urban area of Prague, Czech Republic \cite{prague}. The structure constant of refractive index of air $C_{n}^2$ has been measured throughout months of May to August, and hourly statistics covering the whole measurement period has been used to simulate transmittance values in given conditions. The atmospheric conditions differ significantly throughout the period of 4 months, however turbulence remains weak $\sigma_R^2\approx1$ during the whole duration of channel use. Overall tendencies remain the same - on average the atmosphere is less turbulent during the night, and more during the day: the highest transmittance $\langle\eta\rangle=-1.83 dB$, and lowest fading $Var(\sqrt{\eta})=3.25\times10^{-3}$ is observed around 5pm. Depicted on Fig.\ref{FIG2:2} are the simulation results, that allow to assess whether secure key distribution can be maintained during the whole period of channel use, and how stable can the signal rate be. \par

We analyze the coherent- and the squeezed-state CV QKD protocols (optimized in terms of signal squeezing $V_s$ for the former, and modulation variance $V_m$ for both) in a composite channel with additional fixed losses $\eta_{comb}=-4.5dB$, total excess noise $\epsilon_{+}=1\%$, post-processing efficiency $\beta=95\%$, and finite-size effects were considered for block sizes of $n=10^6$ (dashed lines) and $10^{10}$ (solid lines). The optimized coherent-state protocol (bottom line) with a block size of $n=10^{10}$, on average, most of the operation time cannot reliably distribute signal states between trusted parties and yields secure key only under the best atmospheric conditions, which occur around 5am and 5pm. Under consideration of smaller block size $n=10^{6}$ security of coherent-state QKD cannot be guaranteed under any atmospheric conditions whatsoever. The squeezed-state protocol on the other hand can be successfully implemented for both considered block sizes. While high squeezed states may be costly to generate, even accessible values of $V_s=-3dB$ can significantly improve the performance of the protocol in short atmospheric link with significant additional losses. Provided the finite-size of block to be $n=10^{6}$ the security is threatened only during the worst atmospheric conditions (around 11am) with $Var(\sqrt{\eta})=5.7\times10^{-3}$ and $\langle\eta\rangle=-4.17dB$. Increasing the block size $n$ and/or threshold for squeezing optimization $V_s^{max}$ resolves the issue and completely eliminates downtime of the protocol. Another aspect of the squeezed-state protocol is that  stability of the secure key rate decreases if higher values of squeezing $V_s$ are accessible. In other words, squeezing contribution is more significant in better atmospheric conditions, and this is why the difference between global maxima and minima of the key rate on Fig. \ref{FIG2:2} is greater for the case of $V_s^{max}=-10dB$. 

\begin{figure}[t] 
    \centering
    \includegraphics[width=.99\linewidth]{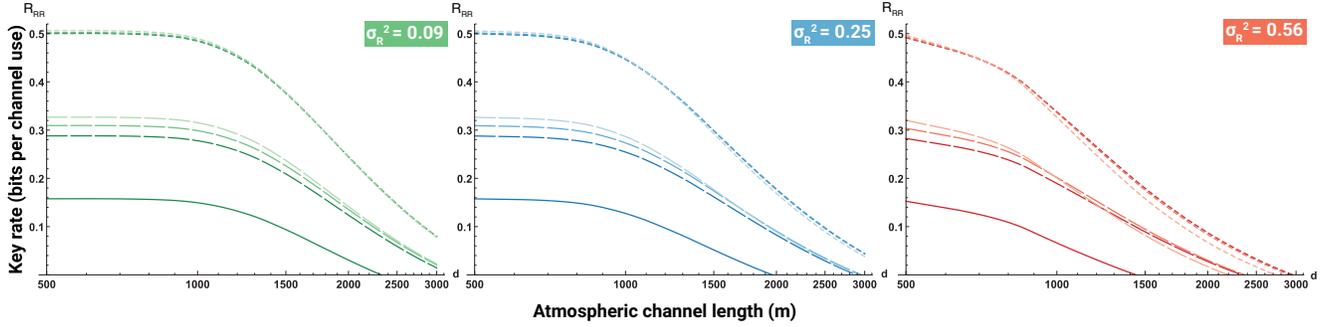}
	\vspace*{-10mm}
    \caption{Secure key rate of the squeezed-state (dashed and dotted lines) and the coherent-state (solid lines) CV QKD protocol on the length (in $m$) of free-space channel in the composite untrusted channel, for various values of Rytov parameter $\sigma_R^2=0.09$ (green), $0.25$ (blue), and $0.56$ (red). Lower values of Rytov parameters yield transmittance distribution $\tau(\eta)$ with smaller variance and higher mean. Reconciliation efficiency $\beta=95\%$, $\eta_{comb}=-4dB$, total excess noise received by Bob $\epsilon_{+}=2.5\%$, modulation $V_m$ is optimized, squeezing is optimized, but limited to experimentally feasible values $V_s^{max}=-3dB$ (long dash lines), and $V_s^{max}=-10dB$ (short dash lines). Finite-size effects have been accounted for block sizes of $n=10^6$. Anti-squeezing noise $V_{AN}=0$ (maximum opacity lines), $+3.1 dB$ (medium opacity lines), and $+10.3dB$ (lowest opacity lines). At short distances ($d\leq500m$) atmospheric links, despite the value of Rytov parameter, have narrow transmittance distributions $\tau(\eta)$, with a mean close to unity, and the performance of the CV QKD protocols is governed by losses in fiber channels $\eta_{comb}$ and total excess noise $\epsilon_{+}$. As atmospheric link length increases, so does the transmittance fluctuations $Var(\sqrt\eta)$, and the need for optimization of the parameters of the CV QKD protocols.}
    \label{FIG3}
\end{figure}

Second example we consider is a CV QKD protocol over short atmospheric links of various length and additional losses before and/or after the link. The atmospheric channels of various lengths have been simulated, using the elliptic-beam model (\ref{Eta}) assuming signal beam wavelength $\lambda=1550nm$, aperture size $a=20mm$, and initial beam-spot radius $W_0=40mm$. The Rytov parameter in all channels $\sigma_R^2<1$, and corresponds to weak turbulence, for which the dominant atmospheric effect is beam wandering. The results of the calculations for the squeezed-state protocol (dashed and dotted lines), and the coherent-state protocol (solid lines) are depicted on Fig. \ref{FIG3}.  All protocols have been optimized in terms of encoding alphabet size $V_m$. The squeezed-state protocol was additionally optimized with regard to signal squeezing, which was limited to attainable values $V_s^{max}=-10dB$ (short dash lines), and $V_s^{max}=-3dB$ (long dash lines). We set post-processing efficiency $\beta=95\%$, and impose significant additional losses in composite channel $\eta_{comb}=-6dB$, as well as excess noise $\epsilon_{+}=2.5\%$. Additionally, we account for finite-size effects, assuming block size of $n=10^6$, and realistic anti-squeezing noise (darkest lines correspond to the protocol with pure states) $V_{AN}=+3.1 dB$ (medium opacity lines), and $V_{AN}=+10.3 dB$ (minimum opacity lines), so that the signal states are initially characterized by $diag[V_s,1/V_s+V_{AN}]$. \par
In fiber channels with fixed attenuation, noise in anti-squeezed quadrature $V_{AN}$ is usually slightly beneficial for trusted parties, since it does not affect mutual information $I_{AB}$ between them, but at the same time reduces the Holevo bound $\chi_{BE}$. However in fading channels this is not the case, as noise in anti-squeezed quadrature, again doesn't alter the mutual information $I_{AB}$, but can increase the Holevo bound $\chi_{BE}$. Despite this the squeezed-state protocol can still significantly outperform coherent-state protocol even under substantial anti-squeezing noise. On the other hand modulation in both ($X$ and $P$) quadratures of coherent-state protocol is a sub-optimal approach if the untrusted channel exhibits significant transmittance fluctuations.  It is certainly beneficial for trusted parties to employ either heterodyne detection, or homodyne detection and modulation of only signal quadrature, with the latter being more advantageous in channels with stronger fluctuations. In our example for channels with Rytov parameter $\sigma_R^2=0.09$, $0.25$, fluctuations of transmittance are low enough so that anti-squeezing noise is actually helpful for the squeezed-state protocol with $V_s^{max}=-3dB$, and does not considerably alter the performance of the protocol with $V_s^{max}=-10dB$. The CV QKD protocols established over the channel with the highest Rytov parameter $\sigma_R^2=0.56$ exhibit on short distances the advantages of anti-squeezing noise, and on longer distances, the noise is conversely more harmful for the CV QKD. The anti-squeezing noise presence elevates the need for squeezing optimization. \par
For very short distances $d\leq500m$ the distinction between channels with different Rytov parameter is insignificant, and secure key rate of the protocols is mainly determined by the excess noise $\epsilon_{+}$ and additional losses $\eta_{comb}$ in such composite channels. The optimal values of signal state squeezing in such regime are equal to maximally permitted for the protocol $V_s^{max}$, while the variance of modulation $V_m$ is mainly limited by efficiency of post-processing algorithms $\beta$. This is of course due to low values of fluctuations of transmittance $Var(\sqrt{\eta})$ in atmospheric channel of such short lengths. \par
As the length of atmospheric channel increases, so does the variance of transmittance $Var(\sqrt{\eta})$, and as consequence the secure key rate starts to drop. However, variance of transmittance $Var(\sqrt{\eta})$ reaches maximum at certain distance (determined by the value of the Rytov parameter $\sigma_R^2$) and then slowly decreases with the distance. In given example for $\sigma_R^2=0.56$ transmittance variance peaks around 1750$m$ at  $Var(\sqrt{\eta})=2.7\times 10^{-3}$, for $\sigma_R^2=0.25$ at around 2000$m$ with $Var(\sqrt{\eta})=1.2\times 10^{-3}$, and for $\sigma_R^2=0.09$ maximum variance $Var(\sqrt{\eta})=4\times 10^{-4}$ is for the 2250$m$ atmospheric channel. Even though the expected fluctuations in simulated channels are low, modulation optimization must be performed to maximize the key rate and reach longer distances for both coherent- and squeezed-state protocols. Squeezing of signal states yields a clear advantage over coherent states, however squeezing optimization is beneficial only for the channels of length 2000$m$ and longer, or for atmospheric channels where turbulence is described by higher values of Rytov parameter. \par
Overall both coherent- and squeezed-state protocols can successfully be implemented over short atmospheric channels even with significant excess noise and additional untrusted losses, but the squeezed states can allow the CV QKD protocol to reach substantially longer secure distances.

\section*{Summary and conclusions}

In the channels with fixed losses the robustness of the CV QKD protocols is unambiguous -- the more losses present in the channel, the less noise the signal can tolerate, and vice versa -- the more noise present in the channel, the less losses the signal can tolerate. The squeezing of the signal states improves the tolerance against both losses and noise in such channels, and the higher levels of squeezing are more advantageous for the protocol, since it will directly translate into considerable improvement in terms of secure key rate\cite{NoiseUF, vmo2, UF10}. In fading channels this is not necessarily the case. Surprisingly, the squeezing is still very beneficial for the security of the CV QKD. However, squeezing of the signal states is advantageous for the protocol but should be optimized depending on the properties of transmittance probability distribution. The effect of transmittance fluctuations is analogous to variance-  and squeezing-  dependent noise, and the more squeezed signal states are used, the more sensitive the resulting protocol can be to transmittance fluctuations in the channel. The presence of fading in the channel limits maximal applicable values of squeezing and modulation variance, and reduces the region of optimal values that allow to maximize the key rate of the protocol. We have proposed the use and shown an  unexpected gain of squeezed-state continuous-variable quantum key distribution protocols in composite untrusted channel, that is the combination of fading atmospheric channel and multiple fiber channels with fixed losses. Our results are compliant with an entanglement-based, as well as prepare-and-measure schemes for the Gaussian states generation, and can be used in all Gaussian CV QKD protocols operated in atmospheric channels. While the coherent-state protocol can only be optimized in terms of displacement of the signal state, it is still a necessary step to reduce the downtime of the protocol. The squeezed-state protocol is sensitive to the fading in untrusted channel, but optimization of squeezing brings considerable benefits, and allows to successfully employ the protocol in greater (comparing to the coherent-state protocol) range of atmospheric conditions, communication distances, and levels of additional losses and noise. The optimization along with the post-selection techniques studied previously \cite{Usenko2012, Derkach2016} can significantly improve the performance of the Gaussian CV QKD protocols in atmospheric links and enable efficient and robust free-space quantum key distribution, fully applicable in daylight conditions. Next step towards this free-space novel quantum key distribution technique is an experimental verification of the functionality of the free-space squeezed-state protocol which will stimulate further theoretical and experimental developments and practical implementations.

\section*{Acknowledgements}
I.D. and V.C.U. acknowledge the project LTC17086 of the INTER-EXCELLENCE program of the Czech Ministry of Education, V.C.U. acknowledges the project 7AMB17DE034 of the Czech Ministry of Education and COST Action CA15220 "QTSpace", I.D. acknowledges the project PrF-2018-010 of Internal Grant Agency at Palacky University.

\section*{Author contributions statement}
V.C.U. and R.F. formulated the idea and the project. I.D. and V.C.U. calculated and simulated the results. R.F. supervised the project. All Authors analyzed and discussed results and contributed to preparation of the manuscript.  

\section*{Additional Information}
\textbf{Competing interests:} The authors declare that they have no competing interests.

\end{document}